\documentclass[review]{elsarticle}

\usepackage{hyperref}

\journal{Journal of \LaTeX\ Templates}

\usepackage{amsfonts,amsmath,amssymb}
\usepackage{graphicx}
\usepackage{mathrsfs}
\usepackage{wasysym}
\usepackage{bm}

\usepackage[svgnames]{xcolor}

\begin{document}

\begin{frontmatter}
\title{Thermoelectric transport and Peltier cooling of cold atomic gases}

\author[adr1]{Charles Grenier}
\author[adr2]{Corinna Kollath}
\author[adr3,adr4,adr5]{Antoine Georges}

\address[adr1]{Laboratoire de Physique, ENS de Lyon, Universit\'e de Lyon, CNRS, 46 all\'ee d’Italie, 69364 Lyon, 
France.}
\address[adr2]{HISKP, University of Bonn, Nussallee 14-16, D-53115 Bonn, Germany.}
\address[adr3]{Coll\`ege de France, 11 place Marcelin Berthelot, 75005 Paris, France.}
\address[adr4]{Centre de Physique Th\'eorique, \'Ecole Polytechnique, CNRS, Universit\'e Paris-Saclay, 91128 Palaiseau, 
France.}
\address[adr5]{DPMC, Universit\'e de Gen\`eve, CH-1211 Geneva, Switzerland.}

\begin{abstract}
This brief review presents the emerging field of mesoscopic physics with cold atoms, with an emphasis
on thermal and `thermoelectric' transport, i.e. coupled transport of particles and entropy. 
We review in particular the comparison between theoretically predicted and experimentally observed thermoelectric 
effects in such systems.
We also show how combining well designed transport properties and evaporative cooling leads to an
equivalent of the Peltier effect with cold atoms, which can be used as a new cooling procedure with improved cooling power 
and efficiency compared to the evaporative cooling currently used in atomic gases. This could lead to a new generation of experiments probing strong correlation 
effects of ultracold fermionic atoms at low temperatures.
\end{abstract}

\begin{keyword}
\texttt{elsarticle.cls}\sep \LaTeX\sep Elsevier \sep template
\MSC[2010] 00-01\sep  99-00
\end{keyword}

\end{frontmatter}

\section{Introduction}

The last fifteen years have established ultracold atomic gases as efficient quantum 
simulators~\cite{feynman1982simulating,RevModPhys.80.885,Bloch:2012aa,lewenstein2012ultracold}.
Using those extremely flexible systems in which geometry, interactions and disorder can be tuned at will, experiments 
on synthetic materials have been performed. These experiments reveal the characteristics of various phase transitions, 
and contribute 
to our understanding of strongly correlated matter.

>From this perspective, it is then quite natural to explore whether this potential for quantum simulation can be 
extended to out-of-equilibrium properties, such as transport, dissipation or thermalization in quantum 
systems. In recent years, numerous experiments and theory proposals have emerged, making cold atom transport a very 
active field~: owing to the flexibility of cold atomic setups, many different configurations have been 
realized~\cite{PhysRevA.85.023623,PhysRevLett.103.150601,ThywiessenQPC,7,8,9,10,17,schreiber2015observation},
which are as many different probes of the out-of-equilibrium properties of quantum systems. In parallel, a whole branch 
of research in cold atoms is now devoted to the realization of circuits in which Bose-Einstein condensates replace 
electrons~\cite{Ramanathan:2011aa,eckel2014hysteresis}, thus founding the field of atomtronics~\cite{18,Atomtronics}.

A similar line of thought is to extend the idea of quantum simulation to two terminal transport setups made of cold 
atoms~\cite{Brantut:2012aa,Krinner,ThywiessenQPC,BrudererPRA}. This has recently brought transport experiments into new 
regimes thanks to the control on interactions~\cite{Stadler:2012kx,Husmann} with Feshbach resonances. This allows one to 
address fundamental questions of mesoscopic physics, such as the interplay between interactions and 
disorder~\cite{PhysRevLett.110.100601,PhysRevLett.115.045302}, or quantized conductance and 
interactions~\cite{BCSconductance}.

Of great interest in this context is thermal transport (i.e the transport of entropy), as well as the combined 
transport of entropy and particles. We shall loosely refer to the latter as `thermoelectric' transport, in line with 
standard condensed matter terminology, notwithstanding that in the present context we are dealing with neutral atoms. 

Combining the flexibility and cleanliness of cold atom systems and the nice transport features of mesoscopic devices is particularly 
exciting, and provides an avenue to improve our understanding of thermal and thermoelectric transport.
Such investigations on clean and controllable systems might also prove helpful to propose systematic guidelines to build 
new materials with interesting transport properties, which can be envisioned for energy saving purposes, for example. In 
return, one may expect that the methods developed in mesoscopic physics might be helpful to tackle some of the 
outstanding problems in cold atom physics, such as cooling.\\

This brief review is organized as follows~: In section \ref{sec:thermo} we discuss thermoelectric effects for cold atoms. 
We introduce our theoretical description and compare the obtained results with experimental data obtained in the quantum optics group at ETH Z\"urich. Then, we show how using a combination of evaporative cooling and well designed transport properties can lead to a better cooling procedure which has a high efficiency and can reach low temperatures, thus, proposing a solution to a long standing problem in the cold atoms community.

\section{Thermoelectricity with neutral particles}
\label{sec:thermo}
This section discusses a theory experiment comparison on thermal and thermoelectric effects concentrating on fermionic atoms~\cite{Brantut08112013}, based on a theory proposal exposed in~\cite{2012arXiv1209.3942G}. A bosonic counterpart of those results exist both on the experimental~\cite{2013arXiv1306.4018H} and 
theoretical~\cite{1367-2630-16-11-113072} side. Thermoelectric effects are fundamental probes for 
materials and investigating them in controlled and flexible systems such as cold atoms is a good opportunity to understand
better the necessary ingredients to improve the thermoelectric figure of merit or power-factor~\cite{macdonaldthermoelectricity}. Here, we discuss the 
model that has been compared to experimental results obtained at ETH. In particular, we show how the magnitude of 
thermoelectric effects and the efficiency of energy conversion can be optimized by controlling the geometry (in the 
ballistic regime) or the disorder strength in the junction. By investigating systematically the 
ballistic-diffusive crossover as a function of disorder, we also illustrate clearly that thermopower and 
conductance measurements are complementary probes of the transport properties of a given system.\\

We point out that the setup is a good test-bed for investigating the thermodynamics of small systems, by 
demonstrating its potential to explore the mechanisms of energy conversion from heat to chemical energy. 
  \subsection{Basics of cold atom thermoelectricity}

    \subsubsection{Model}

\begin{figure}[ht!]
 \centering
 \includegraphics[width=0.8\linewidth]{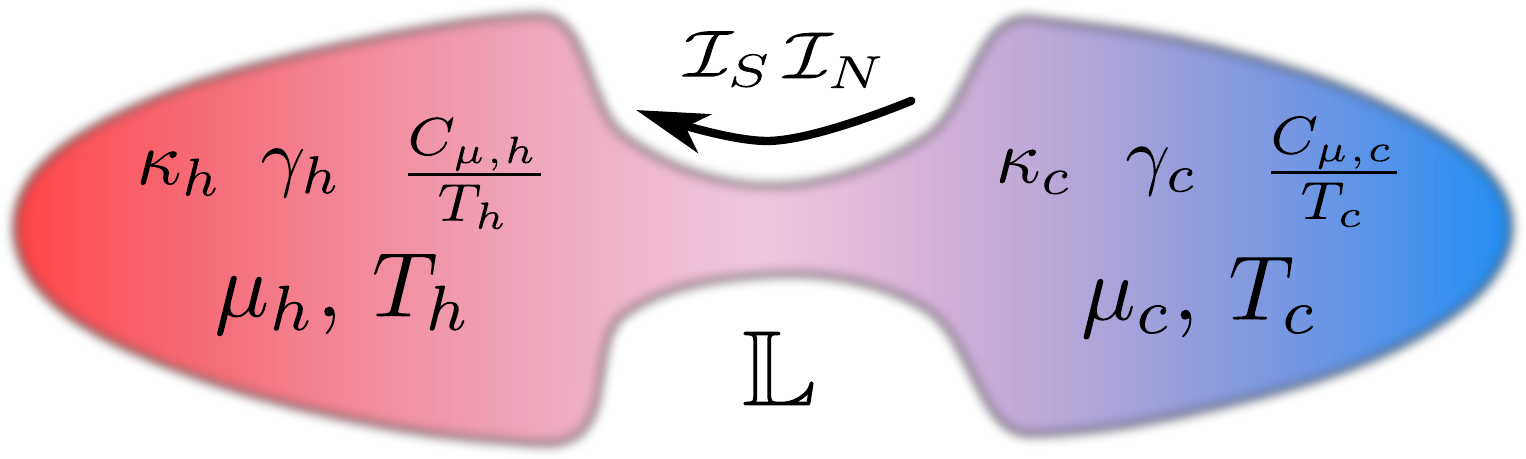}
 \caption{Setup for thermoelectricity (sketch). The two reservoirs ($h$ for hot, $c$ for cold) exchange particles 
 and heat through a junction characterized by its linear response properties represented by the matrix $\mathbb{L}$ 
defined in the main text. The thermodynamic response of each reservoir is represented by the coefficients $\kappa$, 
$\gamma$ and $C_N$, which denote respectively the compressibility, the dilatation coefficient and the heat capacity 
at constant particle number (see text).}
 \label{fig:Thermo_setup}
\end{figure}

The setup under consideration is the two-terminal setup sketched in Fig.~\ref{fig:Thermo_setup}. The setup is largely 
inspired by the ubiquitous Landauer configuration of mesoscopic physics~\cite{datta1997electronic,Imry}. 
A large atomic cloud is divided by a suitable laser beam into two reservoirs which are connected by a narrow channel. 
A detailed description of the setup can be found in~\cite{Brantut:2012aa}

The junction 
between the two reservoirs can be considered as a circuit element having a conductance $G$, a thermal conductance $G_T$ and a 
thermopower $\alpha_{ch}$. In order to get a key signature of thermoelectric effects, the reservoirs are prepared with $N$ atoms each at a temperature $T$. Then, one side is heated up to a temperature $T_h = T + \Delta T_0$, the populations of the two reservoirs 
remaining the same.\\

Cold atom systems are well isolated. This means in particular 
that the total particle number is conserved, implying that current between the reservoirs is only a transient effect and no stationary 
regime will be reached. It is useful to view the above setup as the analogue of a capacitor (each plate being a reservoir) 
connected to a resistor (representing the junction), and the actual experiment as the transient discharge of this 
capacitor.
 
We assume that the temperature and chemical potential biases $\Delta T = T_h - T_c$ and $\Delta \mu = \mu_h - \mu_c$ 
are small with respect to the Fermi temperature $T_F$ and energy $E_F$ (with $T_F = E_F/k_B$) of the entire cloud. 
Therefore, particles and heat flow can be described using linear response and the transport properties of the junction 
by a constant matrix $\mathbb{L}$ of transport coefficients.

Thermoelectric effects originate from a reversible coupling between heat and particle 
flows~\cite{Goldsmid,macdonaldthermoelectricity}. Within linear response, the expressions of the particle and entropy 
currents $\mathcal{I}_N$ and $\mathcal{I}_S$ in terms of the chemical potential and temperature biases $\Delta \mu$ and 
$\Delta T$ are obtained following Onsager's picture of coupled transport 
processes~\cite{PhysRev.37.405,PhysRev.38.2265}~:

\begin{equation}
\label{eq:lin_response}
\left(\begin{array}{c}
        \mathcal{I}_N\\
        \mathcal{I}_S
       \end{array}
\right)
 = 
\frac{d}{dt}\left(\begin{array}{c}
        N_h-N_c\\
        S_h-S_c
       \end{array}
\right)
 =
 -\underline{\mathbb{L}}
\left(\begin{array}{c}
        \Delta \mu\\
        \Delta T
       \end{array}
\right)\,\,\,
\underline{\mathbb{L}} = G\left(
\begin{array}{cc}
  1 & \alpha_{ch}\\
  \alpha_{ch} & \mathcal{L}+\alpha_{ch}^2
\end{array}
\right)\,,
 \end{equation}
where $G$ and $\alpha_{ch}$ are the conductance and thermopower of the junction, respectively, and  
$\mathcal{L} = \frac{G_T}{TG}$ is the Lorenz number of the junction, with $G_T$ the thermal 
conductance~\footnote{Thermoelectic effects are usually described with the heat current instead of the entropy current, 
which we keep here for symmetry reasons.}. Note that, in the present context of neutral particles, $G$, $\alpha_{ch}$, 
$G_T/T$ and $\mathcal{L}$ have the dimensions of $1/h$, $k_B$, $k_B^2/h$ and $k_B^2$, respectively, with $h$ the Planck 
constant and $k_B$ the Boltzmann constant.

For usual ({\it electronic}) condensed matter systems, the (electro)chemical potential and temperature differences at 
the boundaries of the circuit element are imposed through metallic reservoirs. This is a major difference with cold 
atomic systems, which are intrinsically isolated, and are thus described in the microcanonical ensemble. Experimentally,
this feature translates into the application of a particle number and entropy imbalance between the two reservoirs, 
instead of a combined chemical potential and temperature imbalance. Taking this into account requires to relate the 
chemical potential and temperature differences to the population and entropy imbalances at any time. Under the 
assumption of a quasistatic evolution~\footnote{Experimentally, data points are acquired approximately every 100 ms, a 
time longer than the thermalization time in the reservoirs, expected to be around 10 ms.},  $\Delta N$ and $\Delta S$ 
are related to $\Delta \mu$ and $\Delta T$ through the following set of thermodynamic relations~:
\begin{equation}
\label{eq:thermodynamics}
 \left(
 \begin{array}{c}
  \Delta N\\
  \Delta S
 \end{array}
\right)
=
\mathbb{K}
\cdot
\left(
 \begin{array}{c}
  \Delta \mu\\
  \Delta T
 \end{array}
\right)
=\kappa
\left(
\begin{array}{cc}
 1 & \gamma\\
 \gamma &\ell + \gamma^2
\end{array}
\right)\cdot
\left(
 \begin{array}{c}
  \Delta \mu\\
  \Delta T
 \end{array}
\right)\,,
\end{equation}
which define the linear thermodynamic response of the reservoirs. The relations~\eqref{eq:thermodynamics} have been 
cast in a form that is very similar to that of the linear transport equations~\eqref{eq:lin_response}, except that  
thermodynamic coefficients instead of transport coefficients are involved here. In this expression,  
$\kappa = \frac{\partial N}{\partial \mu}|_T$ is the compressibility, 
$\gamma = \frac{1}{\kappa}\cdot\frac{\partial N}{\partial T}|_{\mu}=\frac{1}{\kappa}\cdot\frac{\partial 
S}{\partial\mu}|_{T}$ is a dilatation coefficient measuring the variation of the particle number with 
temperature~\footnote{One also has $\gamma = -\frac{\partial \mu}{\partial T}$ at constant particle number.}, and 
$\ell$ 
is a thermodynamic analogue of the Lorenz number for the reservoirs. It can be easily shown that $\ell = 
\frac{C_N}{T\kappa}$, with $C_N$ the specific heat at constant particle number.

The exact expression of the thermodynamic coefficients is extracted from the equation of state of the gas
in the reservoirs. As for the transport properties of the junction, eq.~\eqref{eq:thermodynamics} assumes that 
the chemical potential and temperature biases are rather small compared to $E_F$ and $T_F$. As discussed in the 
supplementary material of~\cite{Brantut08112013}, this linear response approximation is surprisingly 
robust. In addition, the reference thermodynamic equilibrium state in which the matrices $\mathbb{K}$ and $\mathbb{L}$ 
should be computed is the final one, in which both reservoirs have the same particle number $\bar{N} = 
\frac{N_{tot}}{2}$ and temperature $\bar{T} = T + \frac{\Delta T_0}{2}$, and consequently have the same chemical 
potential $\bar{\mu}$.\\

The thermodynamic coefficients appearing in~\eqref{eq:thermodynamics} are given by the 
following formulae, with the density of states of a harmonically trapped, noninteracting gas~$g (\varepsilon)=
\frac{\varepsilon^2}{4(\hbar\omega_x\hbar\omega_y\hbar\omega_z)}$~:
\begin{eqnarray}
\label{eq:thcoeffres_1}
 \kappa &=& 
\int_0^\infty d\varepsilon\,
g(\varepsilon)\left(-\frac{\partial f}{\partial \varepsilon}\right)\\
 T\gamma\kappa  &=& \int_0^\infty d\varepsilon\,
g(\varepsilon) \left(\varepsilon-\mu\right)\left(-\frac{\partial 
f}{\partial
\varepsilon}\right)
\label{eq:thcoeffres_2}\\
\label{eq:thcoeffres_3}
 \frac{C_N}{T} + \kappa\gamma^2  &=& \int_0^\infty d\varepsilon\,
g(\varepsilon) \left(\varepsilon-\mu\right)^2\left(-\frac{\partial 
f}{\partial \varepsilon}\right)\,,
\end{eqnarray}
where $f(\varepsilon) = \frac{1}{1+e^{\beta(\epsilon-\mu)}}$ is the Fermi-Dirac distribution. The response of the 
reservoirs to the temperature and particle number imbalance is assumed to be independent from the transport properties 
of the junction, in analogy with the junction/reservoir separation in the Landauer picture of mesoscopic transport.

  \subsubsection{Theoretical results and predictions}

Combining the equations~\eqref{eq:lin_response} and~\eqref{eq:thermodynamics} gives the following 
equation ruling the time evolution of the population and temperature imbalances~:
\begin{equation}
 \label{eq:Transport}
\tau_0 \frac{d}{dt}\left(\begin{array}{c}
        \Delta N\\
        \Delta T
       \end{array}
\right)
=
-\left(
\begin{array}{cc}
        1 & \kappa\alpha\\
        \frac{\alpha}{\ell\kappa} & \frac{\mathcal{L}+\alpha^2}{\ell}
       \end{array}
\right)
\cdot
\left(\begin{array}{c}
        \Delta N\\
        \Delta T
       \end{array}
\right)
\end{equation}
where the global timescale $\tau_0 = \kappa/G$ has been identified and measured in~\cite{Brantut:2012aa} and $\alpha = 
\alpha_{ch}-\gamma$ is the total thermopower of the system. This last relation shows in particular that the coupled 
particle and heat transport properties of the total system arise as a competition between the thermal expansion of the 
reservoirs represented by $\gamma$ and the thermopower of the junction represented by $\alpha_{ch}$. Those two 
coefficients represent two different mechanisms for entropy transport : $\gamma$ accounts for the entropy cost 
associated to the displacement of a particle from one reservoir to the other, and $\alpha_{ch}$ is directly identified 
as the amount of entropy carried by each particle entering the junction. Integrating~\eqref{eq:Transport} provides the 
time dependence of the population and temperature differences, with  $\Delta N_0$ and $\Delta T_0$ the initial 
population and temperature imbalance, respectively.

The general solution of the evolution equations (\ref{eq:Transport}), given an initial 
particle and temperature difference $\Delta N_0$ and $\Delta T_0$, reads~\footnote{The expressions provided in the 
supplementary information of \cite{Brantut08112013} contains typographical errors, that have been corrected here}:
\begin{eqnarray}
\label{eq:time_evol_N}
  \Delta N(t) &= \left\lbrace\frac{1}{2}\left[e^{-t/\tau_{-}}+e^{-t/\tau_{+}}\right]
+\left[1+\frac{{\cal L}+\alpha^2}{\ell}\right]
\frac{e^{-t/\tau_{-}}-e^{-t/\tau_{+}}}{2(\lambda_{+}-\lambda_{-})}
\right\rbrace \Delta N_0
+\frac{\alpha\kappa}{\lambda_{+}-\lambda_{-}}\left[e^{-t/\tau_-}-e^{-t/\tau_+}\right]\Delta T_0\\
\label{eq:time_evol_T}
  \Delta T(t) &= \left\lbrace\frac{1}{2}\left[e^{-t/\tau_{-}}+e^{-t/\tau_{+}}\right]
-\left[\frac{{\cal L}+\alpha^2}{\ell}-1\right]
\frac{e^{-t/\tau_{-}}-e^{-t/\tau_{+}}}{2(\lambda_{+}-\lambda_{-})}
\right\rbrace \Delta T_0
+\frac{\alpha}{\ell\kappa(\lambda_{+}-\lambda_{-})}\left[e^{-t/\tau_-}-e^{-t/\tau_+}\right]\Delta N_0
\end{eqnarray}
The inverse time-scales $\tau_\pm^{-1}=\tau_0^{-1}\lambda_\pm$ are given by the eigenvalues of the transport matrix: 
\begin{equation}
 \label{eq:timescales}
 \lambda_\pm = \frac{1}{2}\left(1+\frac{{\cal 
L}+\alpha^2}{\ell}\right)\pm\sqrt{\frac{\alpha^2}{\ell}+\left(\frac{1}{2}-\frac{{\cal L}+\alpha^2}{2\ell}\right)^2}\,.
\end{equation}  
These equations can be viewed as ruling the discharge dynamics of a thermoelectric capacitor. 

The channel is modeled by a linear circuit element at the average temperature $\bar{T}$ and chemical potential 
$\bar{\mu}$. Its linear transport coefficients are given by the following expressions, which have a form very 
similar to those of the thermodynamic coefficients in equations~\eqref{eq:thcoeffres_1},~\eqref{eq:thcoeffres_2} 
and~\eqref{eq:thcoeffres_3}~:
\begin{eqnarray}
\label{eq:conductance}
 G &=& \frac{1}{h} \int_0^\infty d\varepsilon\, \Phi(\varepsilon)
\left(-\frac{\partial f}{\partial \varepsilon}\right)\\
\label{eq:thermopower}
 T\alpha_{ch}G &=& \frac{1}{h}\int_0^\infty d\varepsilon\, 
\Phi(\varepsilon)
\left(\varepsilon-\mu\right)\left(-\frac{\partial f}{\partial 
\varepsilon}\right)\\
\label{eq:th_conductance}
  \frac{G_T}{T}+G\alpha_{ch}^2 &=& \frac{1}{h}\int_0^\infty 
d\varepsilon\,
\Phi(\varepsilon) \left(\varepsilon-\mu\right)^2\left(-\frac{\partial 
f}{\partial
\varepsilon}\right)\,
\end{eqnarray}
where $\Phi(\varepsilon)$ is the transport function of the channel. Note the formal similarity with the above equations 
for the thermodynamic coefficients of the reservoirs, with here the transport function playing the role of the density 
of states. A simple interpretation of $\Phi (\varepsilon)$ is the number of channels available for a particle 
having an energy $\varepsilon$~\cite{datta1997electronic}, since at zero-temperature $G(T=0) = 
\Phi(E_F)/h$. In the case of a single channel, the transport function reduces to the transmission probability as 
a function of energy. More generally, for noninteracting particles of mass $M$ propagating along the $y$-direction and 
harmonically confined in the transverse ($x,z$) direction, the transport function reads~:
\begin{eqnarray}
 \Phi (\varepsilon) & = &
\sum_{n_z=0}^{\infty}\sum_{n_x=0}^{\infty}\int_0^\infty dk_y\,\frac{\hbar k_y}{M}\mathcal{T}(k_y)
\delta\left(\varepsilon-\hbar\omega_x(n_x+1/2)-\hbar\omega_z(n_z+1/2)-\frac{\hbar^2k_y^2}{2M}\right)
\label{eq:transport_fn}\\
{} & = & \sum_{n_z=0}^{\infty}\sum_{n_x=0}^{\infty}
\mathcal{T}(\varepsilon-\hbar\omega_x(n_x+1/2)-\hbar\omega_z(n_z+1/2))
\vartheta(\varepsilon-\hbar\omega_x(n_x+1/2)-\hbar\omega_z(n_z+1/2))\,,
\end{eqnarray}
where $\mathcal{T}$ is the transmission probability\footnote{Strictly speaking, the transmission probability depends 
on momentum along the translationary invariant direction, but it is commonly defined as an energy dependent quantity 
without loss of generality.}. The difference between the various transport regimes is contained in the transmission 
probability $\mathcal{T}$. In~\eqref{eq:transport_fn}, the energy conservation condition states that a particle 
entering the channel will distribute its energy between kinetic (propagation with a certain momentum along $y$) and 
confinement (populating a transverse mode along $x$ and $z$).\\

The energy dependence of the transport function $\Phi$ close to the chemical potential creates a particle-hole 
asymmetry which enhances the value of the thermopower $\alpha_{ch}$. This effect is larger when 
the energy dependence is stronger, as is the case when the conduction regime goes from ballistic to 
diffusive, or when the confinement increases. This can be understood by applying a Sommerfeld expansion to the 
formulae~\eqref{eq:thcoeffres_1},\eqref{eq:thcoeffres_2},\eqref{eq:conductance} and \eqref{eq:thermopower}, giving
the following result for the total thermopower of the combined junction and reservoirs system\footnote{This result is 
known as 
the Mott-Cutler formula for thermopower.}~:
\begin{equation}
\label{eq:MCthermopower}
 \alpha = \frac{\pi^2k_B^2T}{3}\left[ \frac{\Phi'}{\Phi}(E_F) - \frac{g'}{g}(E_F)\right]\,,
\end{equation}
where $E_F$ is the common Fermi energy of the system at equilibrium. The first lesson of this formula is that 
the thermopower vanishes at low $T$, in agreement with the low temperature behaviour of entropy. 
Also, at high temperature when the Fermi function in the expression of the transport coefficients can be replaced by a 
Boltzmann distribution, we see that $\alpha \propto 1/T$. These considerations imply qualitatively that thermoelectric 
effects 
are expected to be maximal around the Fermi temperature.\\
For a ballistic junction, when $E_F \gg \hbar\omega_{i,ch}$, where $\omega_{i,ch}$ are the transverse confinement 
energies in the junction,  
implying that the number of channels is sufficiently large, the total thermopower can be approximated by the following 
expression~:
\begin{equation}
\label{eq:MCBAllistic}
 \alpha_{\textrm{Ballistic}} = \frac{2\pi^2k_B^2T}{3}\left[\frac{1}{E_F-\frac{\hbar \omega_{x,ch}+\hbar 
\omega_{z,ch}}{2}}-\frac{1}{E_F-\frac{\hbar \omega_{x,r}+\hbar \omega_{z,r}}{2}}\right]\,.
\end{equation}
The confinement frequencies $\omega_{i,ch}$ are significantly bigger than those in the reservoirs $\omega_{i,r}$,
implying that the first term (the thermopower of the junction) is dominating in~\eqref{eq:MCBAllistic} over the 
contribution from the reservoirs.\\

For a diffusive junction in which $E_F \gg \hbar\omega_{i,ch}$, the total thermopower reads in the constant 
scattering time approximation~\cite{Ashcroft}~:
\begin{equation}
\label{eq:MCDiffusive}
 \alpha_{\textrm{Diffusive}} = \frac{\pi^2k_B^2T}{3}\left[\frac{5/2}{E_F-\frac{\hbar \omega_{x,ch}+\hbar 
\omega_{z,ch}}{2}}-\frac{2}{E_F-\frac{\hbar \omega_{x,r}+\hbar \omega_{z,r}}{2}}\right]\,,
\end{equation}
showing that thermoelectric effects are expected to be easier to observe, owing to the stronger energy dependence
of the transport function $\Phi$. A systematic comparison of those predictions to experimental data acquired in the 
ballistic-diffusive crossover will be discussed in the following section.\\

\begin{figure}[ht!]
 \centering
 \includegraphics[width=0.95\linewidth]{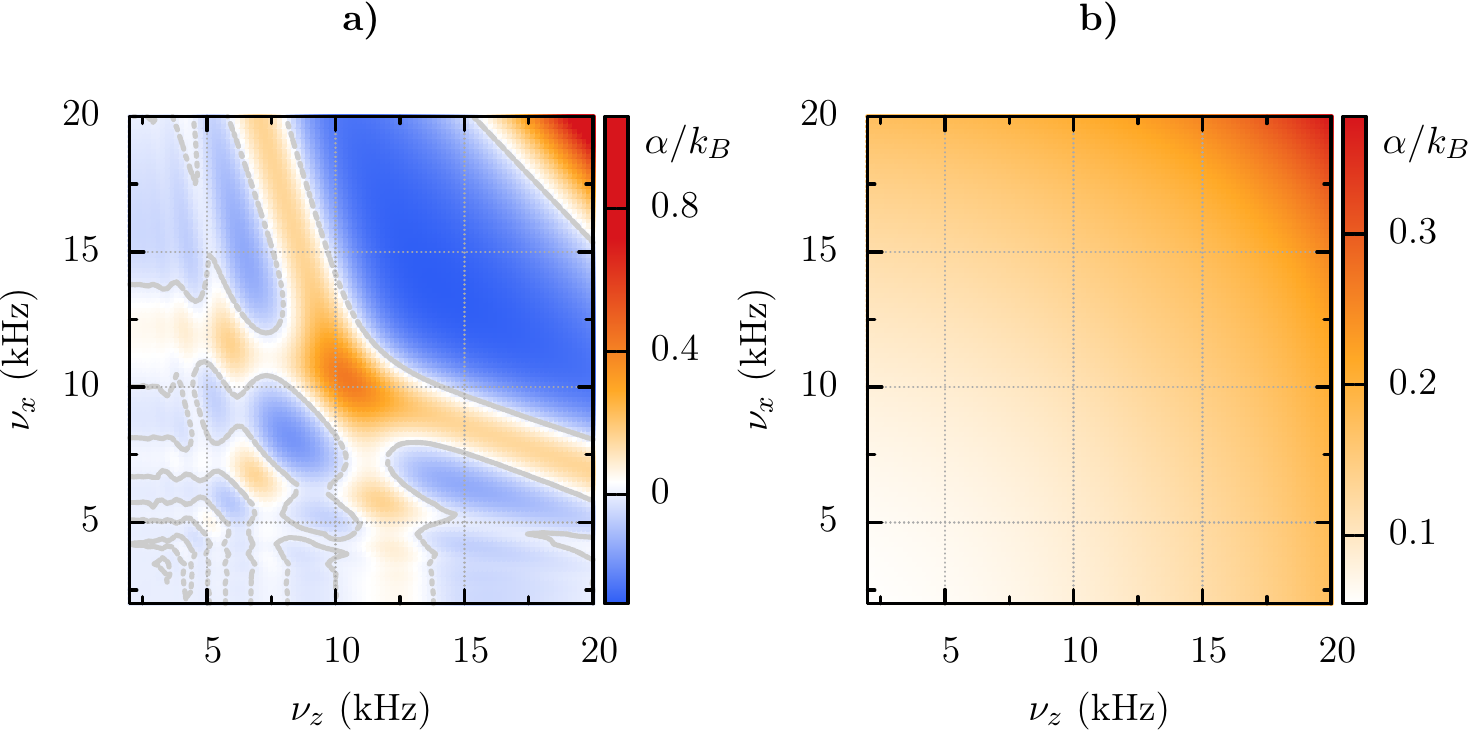}
 \caption{Nanostructuration with cold atoms. The colors indicate the total thermopower $\alpha$ as a function of the 
transverse confinement frequencies $\nu_x$ and $\nu_z$ in the channel, for a fixed Fermi temperature $T_F = 1\mu K$. 
{\bf 
a)} At low temperature $T/T_F = 0.05$ oscillations are visible. Gray lines indicate vanishing total thermopower. {\bf 
b)} At high temperature $T/T_F = 0.3$, the oscillations are washed out.}
  \label{fig:nanostructuration}
\end{figure}

Entering the quantum regime of conduction has also interesting consequences on thermopower, which are known as 
nanostructuration in the context of mesoscopic physics. For a junction with transverse modes having energies 
$\varepsilon_{n_x,n_z}$, the thermopower reads~:
\begin{equation}
 \alpha_{ch} = k_B\frac{\sum_{n_x,n_z\geq 0} 
\left(\xi_{n_x,n_z}-\xi\right)f(\varepsilon_{n_x,n_z})+\log{\left[1+e^{\xi-\xi_{n_x,n_z}}\right]}}{\sum_{n_x,n_z \geq 
0} f(\varepsilon_{n_x,n_z})}\,,
\end{equation}
with $\xi=\beta\mu$ and $\xi_{n_x,n_z} = \beta\varepsilon_{n_x,n_z}$. As depicted in 
Fig.~\ref{fig:nanostructuration}a), this expression for the thermopower leads to oscillations as a function of the 
trapping frequency~: negative values correspond to a reservoir dominated thermoelectric transport, where particles flow 
from the cold side to the hot one, according to the chemical potential difference, and positive to a regime in which 
the 
channel dominates, as in the experiments described below. Similar oscillations having their origin in the successive 
opening of new conduction channels have been predicted and observed in quantum point 
contacts~\cite{molenkamp1991oscillatory}. These oscillations in thermopower as a function of confinement illustrate in 
a very clear way the nanostructuration ideas proposed by Hicks and 
Dresselhaus~\cite{hicks1993effect,hicks1993thermoelectric} 
more than twenty years ago, and they tend to disappear at higher temperature, as shown in 
Fig.~\ref{fig:nanostructuration}b). When $T/T_F$ is sufficiently large, only the overall increase in amplitude remains 
visible, as observed in the experiment discussed in the following section.

  \subsection{Thermoelectric signature}

The forthcoming paragraphs will compare the theoretical predictions obtained by the transport 
equations~\eqref{eq:Evolution_transport1} and~\eqref{eq:Evolution_transport2} to experimental data. The situation we 
consider is the creation of a temperature imbalance between the two reservoirs. Experimentally this has been achieved 
by 
preparing the two reservoirs with the same number of atoms in both reservoirs and then heating one of the reservoirs 
while the junction is closed~\cite{Brantut08112013}. First, we will investigate the transport through a ballistic 
junction. Then, we apply a disordered speckle pattern which changes the properties of the junction to a diffusive 
junction. In particular we will consider thermoelectric manifestations and illustrate the complementarity of resistance 
and thermopower measurements.

\subsubsection{First signature of thermoelectric effects and the ballistic regime}
    
\begin{figure}[ht!]
  \centering
  \includegraphics[width=0.95\linewidth]{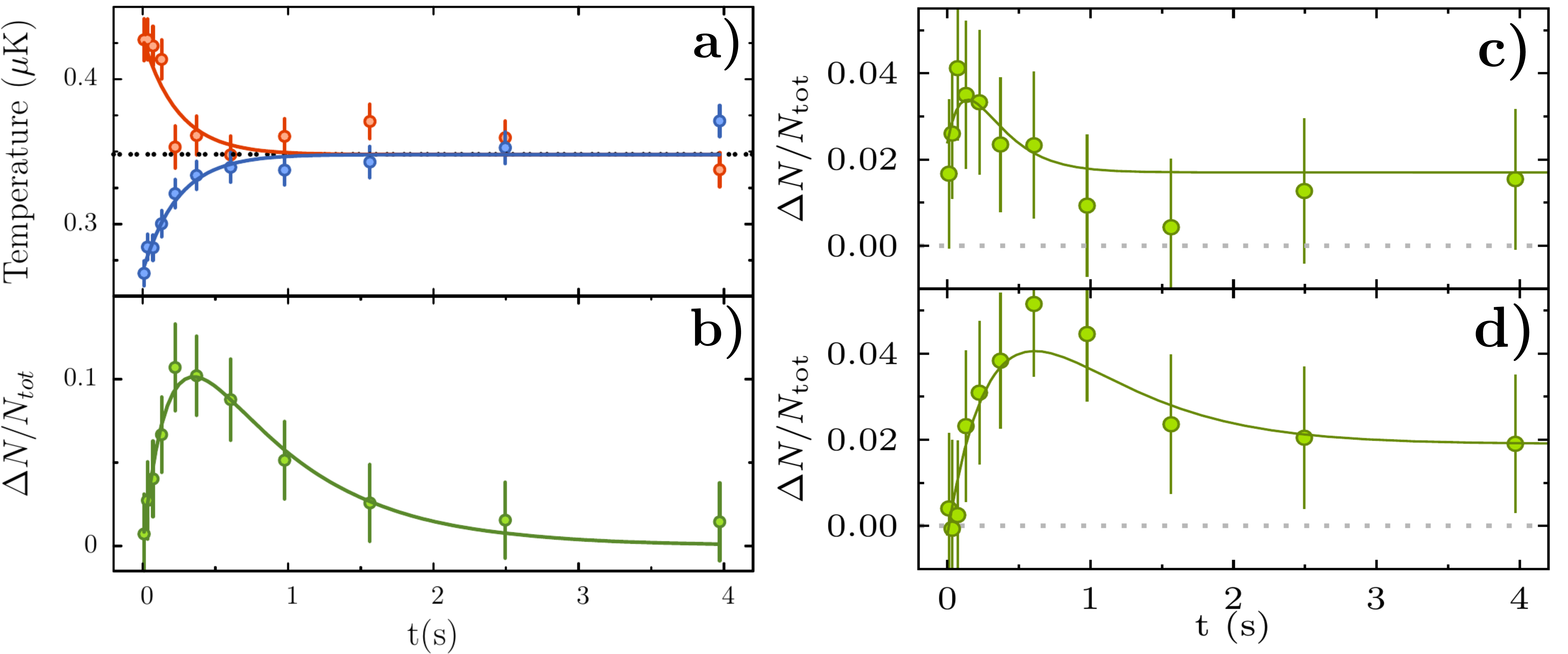}
  \caption{Temperature and population imbalances, adapted from~\cite{Brantut08112013}. {\bf a)} : Temperature 
evolution : $T_h$ (red) and $T_c$ (blue) as a function of time for $\nu_z = 3.5\,$kHz and a disorder of average 
strength  $542\,$nK~(see main text). Dashed line~: $\bar{T}$ at the initial time. The 
average temperature does not evolve significantly with time, in agreement with linear response. {\bf b)} : Population 
relative imbalance $\Delta N/N_{tot}$ as a function of time for a disordered channel in which $\nu_z$ was set to 
$3.5\,$kHz with a disorder of average strength $542\,$nK~(see main text). {\bf c)} and {\bf d)}~: Time evolution of 
$\Delta N/N_{tot}$ a ballistic channel with $\nu_z = 3.5\,$kHz ({\bf c)}) and $\nu_z = 9.3\,$kHz ({\bf d)}).
}
  \label{fig:results_typical_ballistic}
\end{figure}

 Typical time evolutions of the temperature and population imbalances for a ballistic junction are displayed in 
Fig.~\ref{fig:results_typical_ballistic}a) and b), respectively. In these figures the experimental results 
\cite{Brantut08112013}  are compared to the solutions~\eqref{eq:time_evol_N} and~\eqref{eq:time_evol_T} of the 
transport 
equations, both obtained for a Fermi temperature of $T_F \simeq 1\mu K$ and a temperature of $0.35T_F$. 

Two main effects occur. The first effect is the exponentially fast equilibration of the temperatures 
(Fig.~\ref{fig:results_typical_ballistic}a) ) in the two reservoirs which are initially strongly imbalanced. The second 
effect is that even though initially the particle number is almost balanced in the reservoirs, a transient particle 
imbalance is induced. Since the hot reservoir has expanded during the heating process in its trap and is thus less 
dense, naively, one expects a flow from the cold reservoir to the hot reservoir. In other words, a chemical potential 
imbalance in the reservoirs is introduced, since by the heating the chemical potential decreases, which would induce a 
current from the cold to the hot reservoir. However, as one sees in Fig.~(Fig.~\ref{fig:results_typical_ballistic}b) a 
flow from the hot reservoir to the cold reservoir occurs at initial times. This counterintuitive direction of the flow 
is a direct consequence of the thermoelectric coupling in the junction which overwhelms the effect from the reservoirs. 
The particle imbalance reaches a maximum when the typical equilibration timescale of the temperature has been reached. 
Afterwards, a slow equilibration of the particle imbalance takes place to restore thermodynamic equilibrium.

In the ballistic regime, the transverse confinement frequencies can be employed in order to tune the 
transport properties of the junction. The effect of the confinement in the $z$ direction on the thermoelectric response 
is displayed in Fig.~\ref{fig:results_typical_ballistic}(c)-(d). For a larger trapping frequency $\nu_z$ a larger 
induced particle number imbalance can be seen, showing that the maximal thermoelectric response $\mathcal{R}$ increases 
with the trapping frequency $\nu_z$. In both cases the comparison of the theoretical results for the induced particle 
imbalance and the experimental measurements is very good.

The increased thermoelectric response with the larger trapping frequency can be understood from a simple 
estimate~\eqref{eq:MCBAllistic} of the total thermopower. 
Computing the derivative of $\alpha$ with respect to one of the confinement frequencies in the channel gives 
$\frac{\partial\alpha}{\partial\omega_{i,ch}} = \frac{\hbar\pi^2k_B^2T}{3} \frac{1}{\left(E_F-\frac{\hbar 
\omega_{x,ch}+\hbar \omega_{z,ch}}{2}\right)^2}>0$. This implies that the thermoelectric coupling increases with the 
confinement in the junction, in agreement with the experimental observations. Nevertheless, the effect in this 
ballistic regime remains hard to observe, and can barely be distinguished from the error bars.

\subsubsection{Tuning thermoelectric effects : from ballistic to diffusive junctions}

Projecting a disordered potential onto the junction, and thus reaching the diffusive regime, appears as a possibility 
to improve the junction's thermopower. We consider again the situation in which an initial temperature imbalance is 
imprinted onto the junction. 

The microscopic description of the disordered junction is very involved. Thus, we base our theoretical interpretation 
of the ballistic-diffusive crossover on a phenomenological transparency $\mathcal{T}$. This transparency $\mathcal{T}$ 
depends on the energy $\varepsilon$ of incident particles, and on the typical height $\bar{V}$ of the disorder 
potential~:
\begin{equation}
\label{eq:crossover}
 \mathcal{T}(\varepsilon,\bar{V}) = \frac{l(\varepsilon,\bar{V})}{l(\varepsilon,\bar{V})+L} = 
\frac{v(\varepsilon)\tau(\varepsilon,\bar{V})}{v(\varepsilon)\tau(\varepsilon,\bar{V})+L}\,.
\end{equation}
In~\eqref{eq:crossover}, $l$ is the mean free path, $v(\varepsilon) = \frac{\hbar k(\varepsilon)}{M}$ is the velocity 
of the carriers, $L$ is the junction's length and $\tau$ is the scattering time. The expression~\eqref{eq:crossover} 
tends to one when the mean free path $l(\varepsilon,\bar{V})$ becomes larger 
than $L$, as required in the ballistic regime. When the disorder strength becomes large, $l(\varepsilon,\bar{V}) \ll L$,
 the transport function reduces to that obtained in the Drude model describing Ohmic conduction~\cite{Ashcroft}. 
 We assume that the dominant dependence of the scattering time stems from the speckle power with the form $\tau\propto 
\bar{V}^B$ and that its energy dependence can be neglected (a common approximation in solid state systems). In this 
situation the thermopower is independent of the values of the disorder~\cite{Ashcroft}.

\begin{figure}[ht!]
\centering
\includegraphics[width=0.95\linewidth]{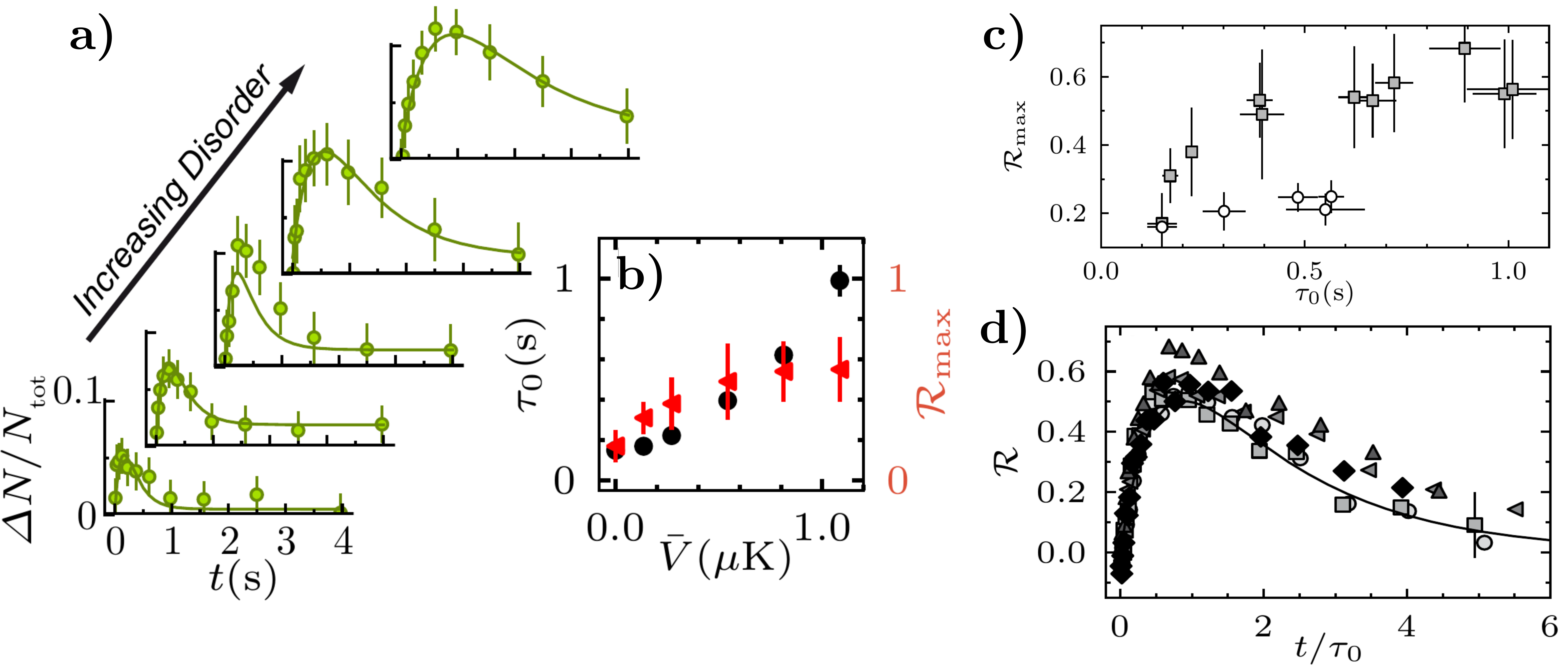}
\caption{Ballistic-diffusive crossover, adapted from~\cite{Brantut08112013}. {\bf a)}: Time evolution of $\Delta 
N/N_{tot}$ for a fixed confinement of $\nu_z = 3.5\,$kHz across the ballistic-diffusive crossover, with increasing 
disorder strength ($\bar{V}=0.14,0.27,0.54,0.81$ and $1.08$\,$\mu$K from bottom to top). Solid lines: theory obtained 
with the transparency in eq.\eqref{eq:crossover}. {\bf b)}: Fitted timescale $\tau_0$ (black circles) and 
$\mathcal{R}_{max}$ (red triangles) as a function of disorder strength for the data set in {\bf a)}. {\bf c)}: Maximal 
response $\mathcal{R}_{\mathrm{max}}$ versus timescale $\tau_0$ for the diffusive (gray squares) and ballistic case 
(open circles). {\bf d)}: Thermoelectric response $\mathcal{R}$  in the regime of strong disorder from $\bar{V} = 
542\,$nK (gray circles) to $1220\,$nK (black diamonds) and fixed $\nu_z = 4.95\,$kHz, in which the time dependence has 
been rescaled by $\tau_0$. Black line: theoretical calculations.}
\label{fig:Result_diffusive}
\end{figure}

The results displayed in Fig.~\ref{fig:Result_diffusive}a) show that the presence of a disorder speckle potential in 
the 
junction has two effects~: First, with increasing disorder strength, the typical transport timescale $\tau_0$ 
increases, since the resistance of the junction increases. The value of $\tau_0$ can been obtained through a fit of the 
population and temperature imbalance with the solution to the evolution equation~\eqref{eq:Transport}. 
Secondly, the disorder leads to an increase of the thermoelectric response directly observed in the time evolution of 
the population imbalance. A measure for this increase is the thermoelectric response 
$\mathcal{R}(t) = \frac{\Delta N(t)/N}{\Delta T_0}$, which removes the initial temperature dependence. A particular 
insight gives the maximal thermoelectric response $\mathcal{R}_{max}$. This quantity is directly proportional to the 
Seebeck coefficient, and thus characterizes the strength of thermoelectric effects.

Fig.~\ref{fig:Result_diffusive}b) summarize this information by comparing the increase of the time-scale $\tau_0$ and 
of 
$R_{max}$ versus the height of the speckle potential. The rise of the time-scale $\tau_0$ and $R_{max}$ imply that the 
resistance and the Seebeck coefficient of the junction increase with the disorder. The resistance essentially 
increases with disorder and confinement, whereas the thermoelectric response first increases and then saturates in the 
diffusive regime. This different behaviour is exhibited in Fig.~\ref{fig:Result_diffusive}c), where $R_{max}$ is shown 
versus the time-sale $\tau_0$. The saturation of the thermoelectric response validates the energy independent 
relaxation 
time approximation. This last feature results, at strong disorder, in a scaling of the time dependent thermoelectric 
response displayed in 
Fig.~\ref{fig:Result_diffusive}d) as a function of the dimensionless time $t/\tau_0$. In addition, the different 
behaviours of $\tau_0$ and $\mathcal{R}_{max}$ clearly show that resistance and thermopower are independent properties 
describing  
the transport in the junction.

    \subsubsection{An ultracold heat engine}

\begin{figure}[ht!]
  \centering
  \includegraphics[width=0.95\linewidth]{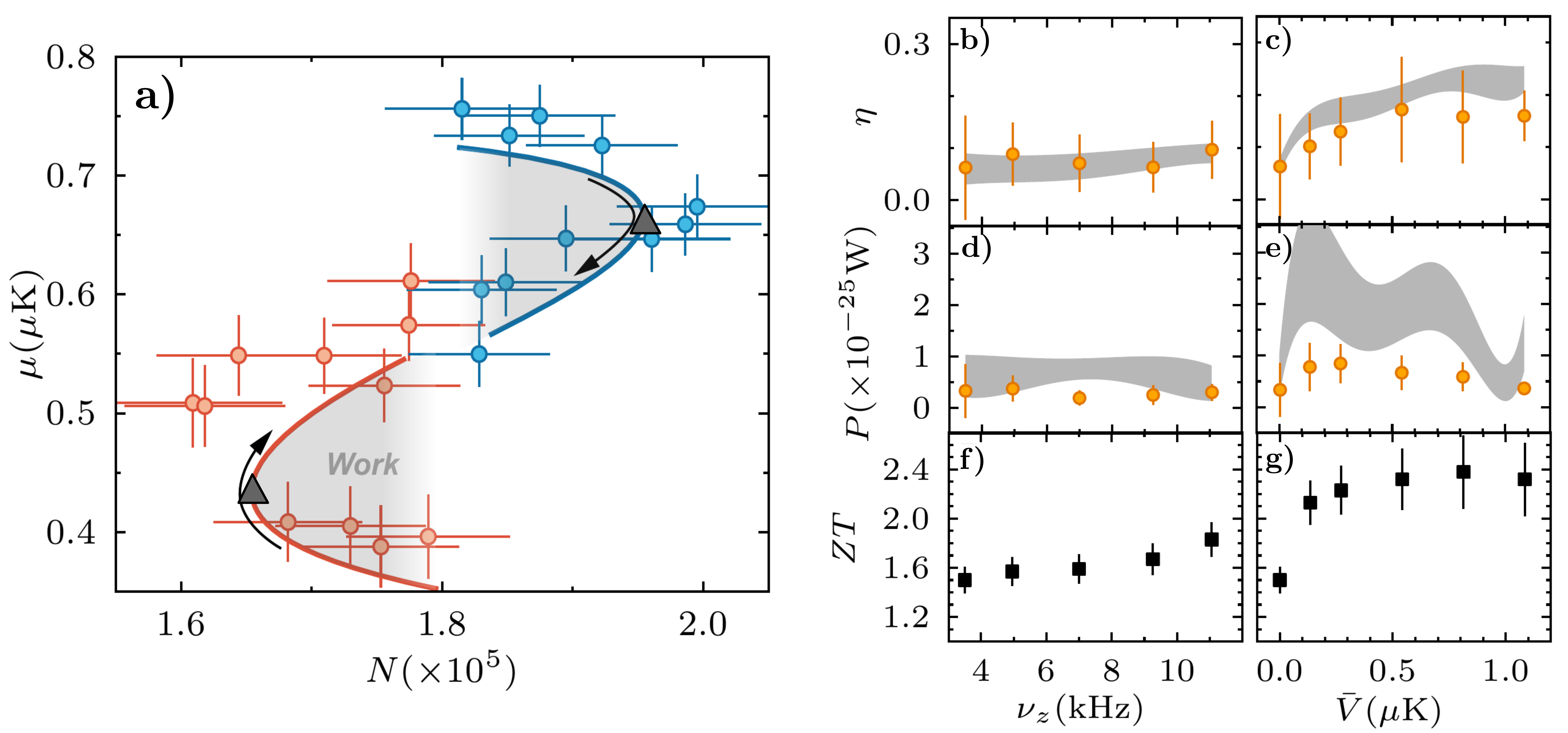}
  \caption{Heat to work conversion, adapted from~\cite{Brantut08112013}. {\bf a)}: Thermodynamic cycle performed by the 
system in the $\mu$-$N$plane for $\nu_z = 3.5\,\mathrm{kHz}$ and $\bar{V}=542\,\mathrm{nK}$. The evolution of the 
hot reservoir is depicted in red and that of the cold one in blue. The solid lines are given by the theory, and
the black arrows indicate the direction of time. The two gray triangles indicate the turning points at which the 
conversion process ends. {\bf b),d)}~: Efficiency, power of the channel in the ballistic case, as a function of 
confinement. {\bf c),e)}~: The same quantities as a function of disorder strength for $\nu_z = 3.5\,$kHz. Orange 
symbols: experiments; grey area: theory. {\bf f),g)}~: Dimensionless figure of merit $ZT$ as a 
function of confinement and disorder for ballistic and diffusive channels respectively.}
  \label{fig:Heat_engine_results}
\end{figure}

>From the thermodynamic point of view, the discussed protocol performs a heat to current conversion from the initial 
temperature
imbalance to the time dependent (transient) particle current. The thermodynamic cycle performed in the $N-\mu$ plane by 
this cold atom heat engine is depicted in Fig.~\ref{fig:Heat_engine_results}a)~: its center signals the final 
equilibrium, where the conversion process ends. The starting points have the same particle number but, owing to the 
initial temperature imbalance, they have different chemical potentials. The latter is responsible for the 'reservoir' 
thermoelectric effect, which will tend to reduce the efficiency of the thermoelectric conversion performed in the 
junction. 
As in the case of a usual heat 
engine~\cite{Callen:Thermodynamics}, the area enclosed by the cycle yields the total work. Strictly speaking, the 
conversion process ends at the turning point of the cycle, shown as dark gray triangles in 
Fig.~\ref{fig:Heat_engine_results}a). Beyond that point the system is essentially performing an equilibration of the 
particle number 
to restore thermodynamic equilibrium and behaves more like a discharging capacitor.\\

All the effective transport coefficients are ratios that depend only on the variable $\frac{\mu}{k_BT}$ or equivalently 
on $T/T_F$. The solutions to the evolution equations~\eqref{eq:Transport} also contain the entropy imbalance and the 
chemical potential difference as a function of time, which allows one to compute the efficiency of the thermoelectric 
process, defined as the ratio between the output chemical work performed during the whole cycle (including the decrease 
in the second stage) and the amount of entropy generated during the evolution. The expression for those two quantities 
is directly extracted from the force-flux formulation of out-of-equilibrium 
thermodynamics~\cite{PhysRev.38.2265,PhysRev.37.405,Callen:Thermodynamics}~:
\begin{equation}
 \label{eq:efficiency_result}
 \eta \equiv \frac{\text{Work}}{\text{Dissipated heat}}= -\frac{\int (\dot{N_c}-\dot{N_h})\cdot(\mu_c-\mu_h)dt}{\int 
(\dot{S_c}-\dot{S_h})\cdot(T_c-T_h)dt} = \frac{-\alpha\alpha_r}{\ell+L+\alpha^2-\alpha\alpha_r}\,.
\end{equation}

The efficiency of the heat to work conversion is depicted in  Fig.~\ref{fig:Heat_engine_results}b)-c) 
in the ballistic and diffusive regime, respectively. The corresponding output powers are displayed in 
Fig.~\ref{fig:Heat_engine_results}d)-e). They show that the efficiency is an increasing function of thermopower, as 
expected, and that the evolution of the efficiency is identical to that of the thermoelectric figure of merit $ZT$ 
plotted in Fig.~\ref{fig:Heat_engine_results}f)-g). As expected from thermodynamic considerations, the slow 
dynamics induced by a strong disorder implies a high efficiency, at the price of low power as seen in 
Fig.~\ref{fig:Heat_engine_results}d) and e). This compromise between efficiency and output power is at the heart of the 
discussion of thermoelectric energy conversion~\cite{CurzonAlborn}, and it is clear that the characterization of the 
thermoelectric performance of a material should be based on both the thermoelectric figure of merit and the power 
factor $G\alpha^2$.\\

This theory-experiment comparison assesses the potential of cold atoms for emulating thermal transport properties. From 
a broader perspective, those results also show that the quantum simulation trend in cold atoms is also valid for 
transport, and could then provide clues on the path to the realization of new materials with interesting energy saving 
properties. From the cold atoms perspective, this study of thermoelectric effects opens a way to the control over heat 
transport with well designed transport properties, which we proposed to use as a mean to overcome fundamental 
limitations in cooling processes.

\section{Making cold atoms cooler}

Low temperatures have been a long standing quest in the cold atom community, especially for 
fermionic gases. The combination of laser cooling and evaporative cooling~\cite{Ket1,PethickSmith}
has proved very efficient to reach temperatures sufficiently low to get quantum degenerate gases
with entropy per particle as low as $T/T_F\approx 0.1$~\cite{0034-4885-74-5-054401}. Nevertheless,
this typical value is still too high to rely on cold atoms to emulate many interesting quantum phases such as 
fractional 
quantum Hall
systems~\cite{CooperReview}, or to investigate the antiferromagnetic order in the Hubbard
model~\cite{0034-4885-74-5-054401,PhysRevLett.104.180401,RevModPhys.80.885}.\\

In~\cite{Peltier_CA}, we have proposed a cooling scheme for fermionic quantum gases, based on the principles of the 
Peltier effect, combined with evaporative cooling. We have shown that both a significantly lower entropy per particle 
and a faster cooling rate can be achieved compared to using only evaporative cooling.

  \subsection{A cold atom Peltier module}

\begin{figure}[ht!]
  \includegraphics[width=0.95\linewidth]{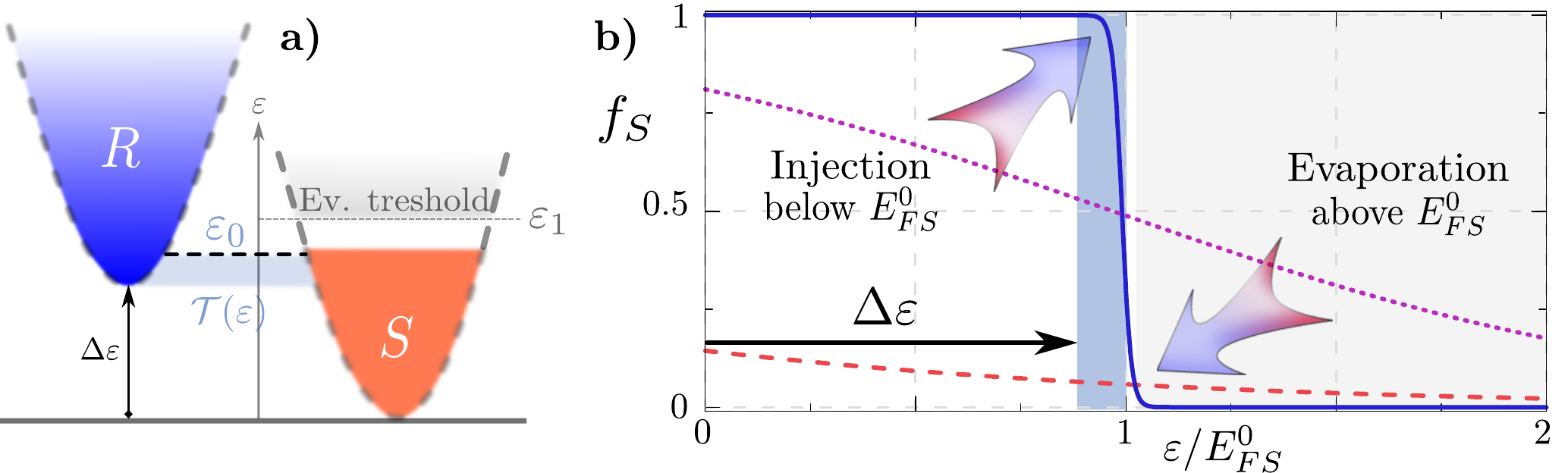}
\caption{Principle of Peltier cooling for cold atoms. {\bf a)} Adapted from~\cite{Peltier_CA}. Sketch of the 
proposed Peltier cooling scheme: atoms are injected from deep energy levels of the reservoir cloud ($R$) into the 
system cloud ($S$) just below the Fermi level $E_{F,S}^{(0)}$ through a channel with an energy-dependent transmission 
$\mathcal{T}(\varepsilon)$. Additionally, the system is submitted to evaporative cooling with a fixed evaporation 
threshold $\varepsilon_1$ located above the Fermi level, removing high energy particles, as indicated in the grey area 
above $S$. {\bf b)} Evolution of the Fermi distribution of the system at three stages during the cooling process: 
initial (dashed red curve, $T_S \approx T_{FS}$), intermediate (purple dotted curve, $T_S = 0.3 T_{FS}$) and final 
(solid blue curve, $T_S=0.02T_{FS}$). The evolution, indicated by arrows, is calculated  for $\varepsilon_1 = 1.05 
E^0_{FS}$, $\gamma_{ev}\tau_0 = 1/16$, $\varepsilon_0 = 0.99E^0_{FS}$, $\Delta \varepsilon = 0.96 E^0_{FS}$ and 
$E^0_{FS} = 0.25 E^0_{FR}$, for an ideal sharp transmission (see below). The blue and grey shaded regions indicate the 
injection and evaporation energy  windows, respectively.}
 \label{fig:Peltier_Setup}
\end{figure}

Our proposal is inspired by the two terminal setup realized at ETH~\cite{Brantut:2012aa} discussed in the previous 
section. The proposed setup is displayed on Fig.~\ref{fig:Peltier_Setup}a). The idea is to cool down one of the 
two clouds (the system $S$) with a technique based on thermoelectric 
effects~\cite{Brantut08112013,2012arXiv1209.3942G,2013arXiv1306.4018H,PhysRevLett.109.084501, papoular2014fast,
Sidorenkov:2013aa,Kim:2012aa,PhysRevA.85.063613,Karpiuk:2012aa} appearing at the junction between $S$ and the second 
cloud (the reservoir $R$), which will ultimately be separated from the system. As its counterpart discussed in the 
previous section, the Peltier effect is a reversible thermoelectric phenomenon.\\
Both the reservoir $R$ and the system $S$ are prepared in harmonic traps, and their initial Fermi energies are 
$E^0_{F,R/S}=h\bar{\nu}(3N_{R/S})^{1/3}$, with $\bar{\nu}$ the average trapping frequency and $N_R$, $N_S$ the atom 
numbers. The reservoir and the system differ in their populations :  $N_R>N_S$ (implying in particular that the ratio
$T_R/T_{F,R}$ is lower than $T_S/T_{F,R}$ in the reservoir). Additionally, the 
lowest energy state of the reservoirs $R$ is 
offset by $\Delta\varepsilon\geq 0$ against the lowest energy state of the system $S$. A 'junction' with an energy 
dependent transmission connects the reservoir and the system in order to allow particle exchange in between the two. 
Within this setup, two different processes can lead to cooling. The first process is the usual evaporative cooling of 
the system in which hot atoms above the Fermi surface are removed from the gas. The second process is the filling in of 
holes below the Fermi surface of the system via the energy resolved junction using atoms from the reservoir. 

Our theoretical description of the cooling process relies on coupled rate 
equations~\cite{PhysRevA.53.381,DavisMewesKetterle,metcalf1999laser,Ket1} for the distribution functions $f_S$ and 
$f_R$. The evolution is driven by the combined action of the junction between the reservoir and the system 
and the evaporation (which acts on the system only). The evolution of the thermodynamic quantities (particle number, 
energy and entropy in our case) in the two harmonic traps is quasistatic, provided that the thermalization is fast 
enough. This quite crucial assumption has been experimentally verified in the two terminal configuration of ETH, even 
in 
the case where many conduction channels are simultaneously open. Under this assumption, the particle current leaving 
the reservoir, given by the Landauer formula, can be directly related to the variation of the particle number in the 
reservoirs, to give the following coupled evolution of the two distribution functions~: 
\begin{eqnarray}
 g_R(\varepsilon)\frac{df_R(\varepsilon)}{dt} & = & 
-\frac{\mathcal{T}(\varepsilon)}{h}\left[f_R-f_S\right](\varepsilon)\label{eq:Evolution_transport1}\\
 g_S(\varepsilon)\frac{df_S(\varepsilon)}{dt} & = & \frac{\mathcal{T}(\varepsilon)}{h}\left[ 
f_R-f_S\right](\varepsilon) 
- \Gamma_{ev}(\varepsilon)g_S(\varepsilon)f_S(\varepsilon)\,,
 \label{eq:Evolution_transport2}
\end{eqnarray}
where $g_{S,R}$ are the densities of states in the system and reservoir. Unless specified, the number of transport 
channels is one, such that the transport function representing the connection between the system and reservoir reduces 
to the energy dependent transmission $\mathcal{T}(\varepsilon)$, which we propose to tune to improve the cooling power. 
Let us note that since $\mathcal{T}(\varepsilon)$ is dimensionless, the typical time-scale that rules the 
time-evolution 
in these equations is $\tau_0 = hg_S(E^0_{FS})=h\left(E^0_{FS}\right)^2/3(h\bar{\nu})^3$, in agreement with the 
relevant 
time-scale for the particle transport identified previously~\cite{Brantut:2012aa,Brantut08112013}. The effect of 
evaporation, acting on the system only, has been included as a leak of high energy particles above a fixed energy 
threshold $\varepsilon_1$, with an energy independent rate $\Gamma_{ev}(\varepsilon) = 
\gamma_{ev}\vartheta(\varepsilon-\varepsilon_1)$~\cite{Ket1}, with $\vartheta$ the Heaviside step function. Excluding 
the trapping geometry, the parameters at hand include the energy bias $\Delta \varepsilon$, the typical transmission 
energy $\varepsilon_0$ and the evaporation threshold $\varepsilon_1$. In~\cite{Peltier_CA}, we performed a systematic 
exploration of the parameter space for different transmissions, monitoring the evolution of the entropy per particle, 
particle number and energy as a function of time.

  \subsection{Fast and efficient cooling with Peltier and evaporation}
 
A good benchmark for a cooling protocol relies on two criteria. The first one is to reach the best cooling efficiency 
in 
the sense of thermodynamics~: in analogy with usual cooling processes~\cite{Callen:Thermodynamics}, this means 
reaching the lowest possible entropy per particle. However, in practice one needs to take into account the speed at 
which the cooling operation is performed, since in experiments other destructive processes may be present such as 
heating processes by the interaction with the applied trapping lasers or scattering with the thermal background gas.
Therefore, it is important to make sure that the {\it rate} at which atoms are cooled is high enough to overcome the 
heating rate imposed by the experimental setup.  Thus, a good cooling protocol must be powerful and thermodynamically 
efficient at the same time. Consequently, we have calculated the cooling rate defined as the amount of entropy per 
particle $s$ removed per unit time~:
\begin{equation}
 \label{eq:rate}
 \eta\tau_0 = \frac{ds}{dt}\,.
\end{equation}

\begin{figure}[ht!]
 \centering
 \includegraphics[width=0.95\linewidth]{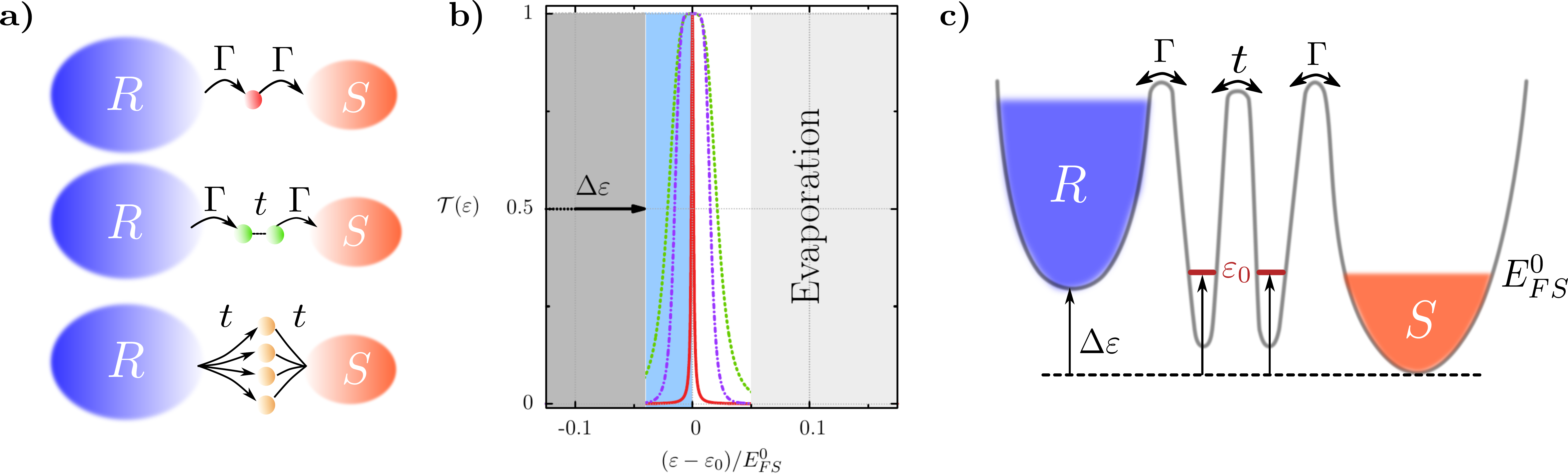}
 \caption{{\bf a)} : Sketches of the different propositions to realize the desired energy dependent transmission 
functions. Top (red dot) : a single resonant level, with a transmission given by~\eqref{eq:1RL}. Middle (green dots) : 
Two resonant levels in series connected by a tunnelling probability, generating a transmission given 
by~\eqref{eq:2RL}, and Bottom (orange dots) : Several resonant levels in parallel. {\bf b)} : Adapted from 
~\cite{Peltier_CA}. The corresponding energy-dependent transmission coefficients, plotted for parameters identical to 
those in Fig.~\ref{fig:Rates}. The light grey area indicates states above $\varepsilon_1$, which are subject to 
evaporation, while dark grey one indicates those below $\Delta \varepsilon$, which do not participate to transport. The 
light blue area corresponds to the ideal box-like transmission, bounded by the band bottom and $\varepsilon_0$. The 
transmission for three resonant levels in series is shown in purple for comparison with that of the two levels in 
series. {\bf c)} : Sketch  of the energy levels for two resonant levels in series.}
 \label{fig:Realizations}
\end{figure}

In the context of solid state devices an energy resolved transmission of a resonant level (\eqref{eq:1RL} below) has 
been shown~\cite{Mahan23071996} to enhances thermopower, and to improve drastically the efficiency of the heat to 
particle current conversion. Nevertheless, a good cooling efficiency does not ensure a good cooling power, which 
depends 
on the power factor $\alpha^2G$. The study performed in~\cite{PhysRevLett.112.130601} showed that optimizing the power 
factor instead of the thermoelectric figure of merit $ZT$ actually leads to a box-like transmission (\eqref{eq:box} 
below), which can be mimicked with resonant levels in series (\eqref{eq:2RL} below). Optimizing the power factor 
instead 
of the figure of merit is equivalent to optimize power instead of efficiency. Starting from these considerations, we 
have compared the efficiency and cooling power of the following three transmissions, which are also shown in 
Fig.~\ref{fig:Realizations}, together with possible experimental implementations~:
\begin{eqnarray}
\label{eq:1RL} \mathcal{T}_{1RL}(\varepsilon) &=& \frac{\Gamma^2}{(\varepsilon-\varepsilon_0)^2+\Gamma^2}\\
\label{eq:2RL} \mathcal{T}_{2RL}(\varepsilon) 
&=& \frac{\Gamma^2t^2}{\left[(\varepsilon-\varepsilon_0)^2+\Gamma^2/4-t^2\right]^2+\Gamma^2t^2}\\
\label{eq:box} \mathcal{T}_{box} & = & \vartheta(\varepsilon-\varepsilon_0)\,.
\end{eqnarray}

\begin{figure}[ht!]
 \centering
 \includegraphics[width=0.95\linewidth]{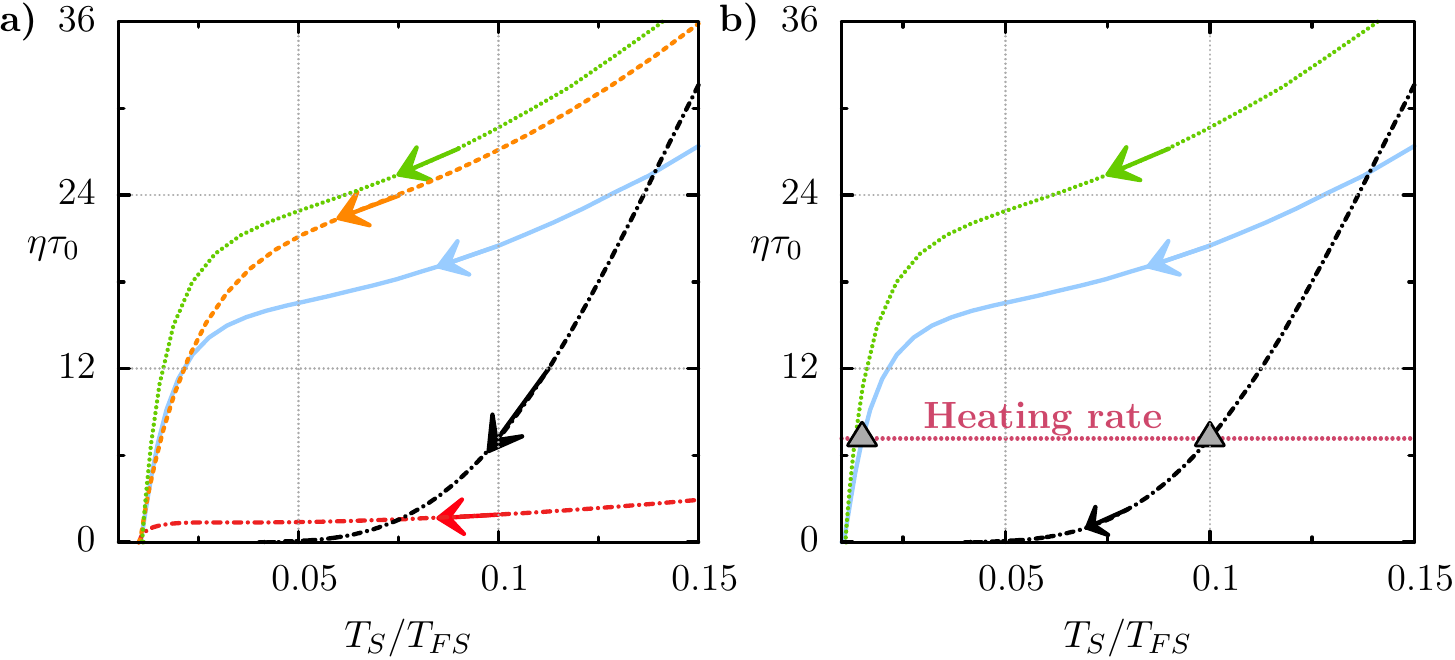}
 \caption{Adapted from~\cite{Peltier_CA}. {\bf a)} : Dimensionless cooling rate $\eta(t)\tau_0$ as a function of 
$T_S/T_{FS}$, for  $\Delta\varepsilon = 0.96E^0_{FS}$ and various transmissions centered at $\varepsilon_0 = 0.99 
E^0_{FS}$ , and $E^0_{FS}/E^0_{FR}=1/4$, $\varepsilon_1 = 1.05E^0_{FS}$, 
$\gamma_{ev}\tau_0 = 15$,  $\Delta \varepsilon = 0.96E^0_{FS}$: 
The (red) dot-dashed and (orange) dashed curve correspond to a single and 100 parallel resonant level(s), respectively, 
with $\Gamma = 1\cdot 10^{-3} E^0_{FS}$. The (green) dotted curve is for two resonant levels in series of width $\Gamma 
= 0.03 E^0_{FS}$ and the (light blue) solid curve is for an ideal box transmission. (Black) dashed-dotted curve shows 
the evaporative cooling only. {\bf b)} :  Dimensionless cooling rate as a function of the entropy per particle, for the 
same parameter values as in {\bf a)}. Arrows indicate the direction of the time evolution. The horizontal (red) dotted 
line indicates a typical heating rate limiting these cooling processes. The gray triangles signal the end of the 
cooling process, defined as the intersection of the curves with the horizontal line accounting for the 
heating rate.}
\label{fig:Rates}
\end{figure}

The cooling rates obtained for the different transmissions are depicted in Fig.~\ref{fig:Rates}a). They show in 
particular that the resonant levels in series or in parallel satisfy both criteria : they offer a good cooling rate, 
and 
they allow to obtain a rather low final entropy per particle. As expected, the results for the single resonant level 
are 
pretty good in terms of the final entropy per particle, but the associated rate is rather poor, since the current 
passing through the single level cannot compensate efficiently the evaporation losses. Also, the result obtained with 
the resonant 
levels in parallel is even better than that obtained with the ideal box transmission, which can be attributed to the 
extra current going through the tails of the transmission function, which makes it more efficient at the initial 
cooling 
stage. Fig.~\ref{fig:Rates}b) displays the cooling rate for the ideal box transmission and for single evaporation, 
together with a phenomenological heating rate, which signals the end of the cooling process, chosen such that 
evaporative cooling allows one to obtain a temperature of $0.1T_F$ approximately. Under the same conditions, the 
Peltier cooling scheme allows to reach temperatures as low as $0.015T_F$.

\begin{figure}[ht!]
 \centering
 \includegraphics[width=0.95\linewidth]{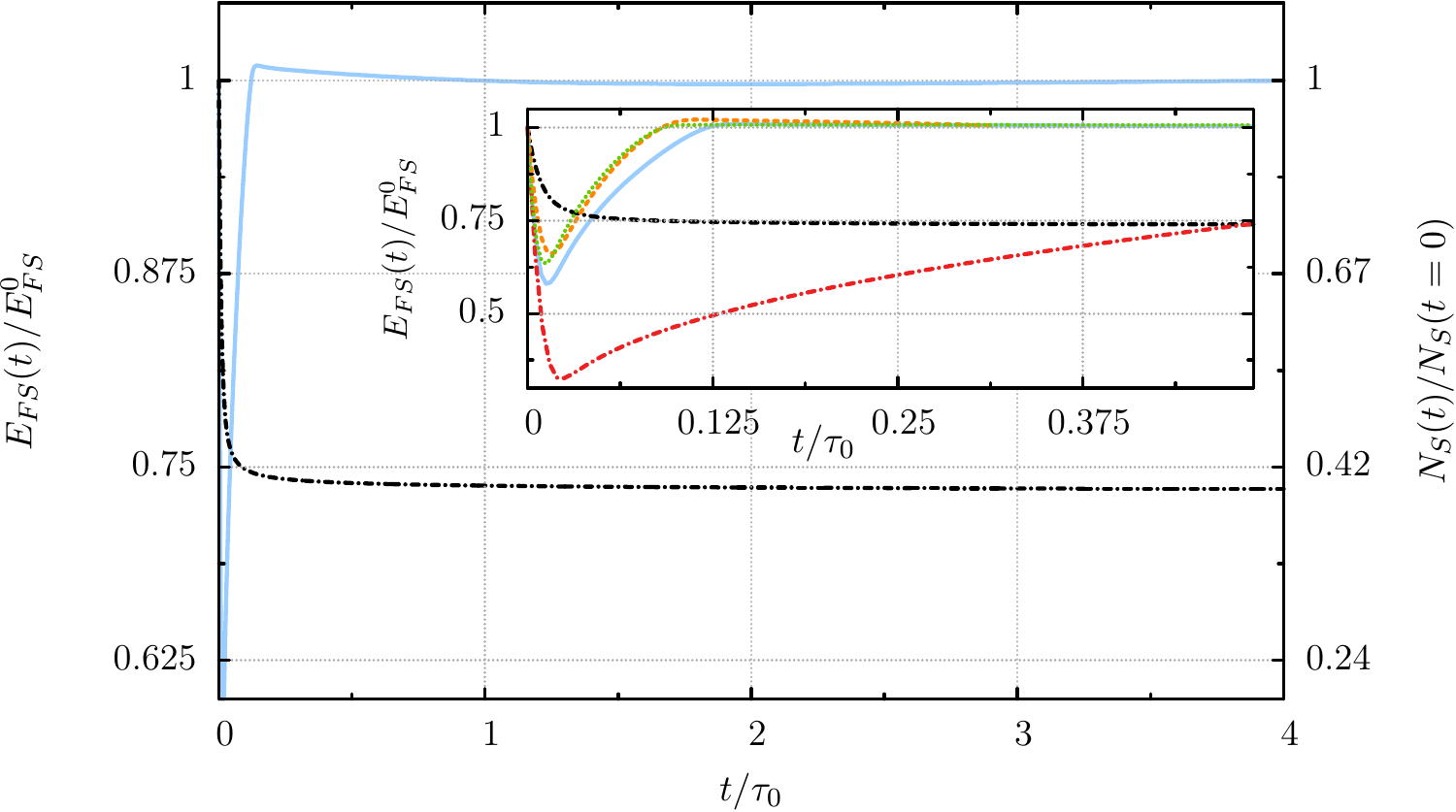}
 \caption{Adapted from~\cite{Peltier_CA}: The system's Fermi energy $E_{FS}(t)$ (left axis) and particle number $N_S(t)$ 
(right axis). The solid light blue curve is for the Peltier cooling with $E^0_{FS}/E^0_{FR}=1/4$ and $\varepsilon_1 = 
1.05E^0_{FS}$, $\gamma_{ev}\tau_0 = 15$, $\varepsilon_0 = 0.99E^0_{FS}$, $\Delta \varepsilon = 0.96E^0_{FS}$. The 
dashed-dotted black curve is for evaporative cooling only, with an initial particle number $N=N_S+N_R$ and 
$\varepsilon_1 = 1.05E_{F}$. {\bf Inset~:} Comparison between the different realizations at short times. The parameters 
are the same as in Fig.~\ref{fig:Rates}.}
\label{fig:EFandN}
\end{figure}

In addition, the results for the evolution of the particle number are represented in Fig.~\ref{fig:EFandN} for an ideal 
box-like transmission, showing that, in contrast to evaporative cooling, our cooling proposal is not accompanied by 
particle losses. The losses due to evaporation are overcompensated by particles from the reservoirs, in the sense that 
we propose to trade high energy 
particles for low energy ones~: with this point of view, our cooling scheme can be understood as the simultaneous 
evaporation of particles and holes at fixed Fermi energy, thus performing a control on entropy transport,
leading to a rectification of the distribution function in the reservoirs, as displayed in Fig.~\ref{fig:Peltier_Setup}. 
The 
inset of Fig.~\ref{fig:EFandN} confirms that the poor results obtained with the single resonant level come from the 
difficulties to compensate for the particle losses. The resonant levels in parallel and those in series perform 
similarly to the ideal box like transmission.

\section{Conclusion and perspectives}

We have demonstrated in~\cite{Brantut08112013} that thermal and thermoelectric transport measurements can be performed 
in cold atomic gases, and that they are sensitive observables, complementary to a conductance measurement.\\ 
Elaborating on this theory-experiment confrontation, we have proposed~\cite{Peltier_CA} a fast and efficient cooling 
scheme for fermionic gases based on evaporation and energy-selective injection of particles. This proposal can be 
readily implemented with state of the art projection techniques~\cite{1367-2630-13-4-043007}, and appears as a plausible 
answer to a long standing goal of low temperatures with cold fermionic gases.\\

Owing to the control on interactions via Feshbach resonances, the experimental technique developed 
in~\cite{Brantut08112013} can be generalized to strongly interacting systems where thermoelectric properties are of 
fundamental interest~\cite{Behnia:2004aa,Zhang:2011aa,Micklitz:2012aa}. Entering the correlated regime would 
make the setup an interesting controllable testbed for theories on the transport properties of strongly interacting 
systems~\cite{PhysRevB.13.647}, and it would be of great interest to extend the existing results on particle 
transport~\cite{Husmann,BCSconductance} to thermal and thermoelectric properties, which appear as efficient probes of 
the excitations of a system, as opposed to the particle current which contains contributions stemming also from the 
ground state. Thus, extending both the theoretical and experimental results presented in the first part to the strongly 
interacting limit of the Hubbard model would be a milestone to the quantum simulation of purely out-of-equilibrium 
properties of strongly correlated models. As an example, investigating the mechanisms responsible for energy exchanges 
in the strongly interacting, insulating phase of the Hubbard model allowing for the relaxation of a temperature 
imbalance would give an insight on genuine many-body effects out of the ground state. Characterizing those mechanisms 
remains a challenging question, both theoretically and experimentally, and any opportunity to investigate them should be 
taken. Along the same lines, exploring attractive interactions would reveal the physics of excitations of strongly 
correlated superfluids, and draw a relation between second sound (and more generally thermomechanical effects) and 
thermoelectricity as entropy conveyors.\\
The recent emergence of periodic driving techniques in cold atomic systems~\cite{jotzu2014experimental} also offer 
interesting perspectives to simulate AC thermopower.\\

{\it Acknowledgments}\\

All the results reviewed here have benefited from valuable discussions with J.-P.~Brantut,  
M. B\"uttiker, T.~Esslinger, S.~Krinner, D.~Stadler, J.~Meineke, H. Moritz, D.~Papoular, 
J.~L.~Pichard, B.~Sothmann, S.~Stringari and R.~S.~Whitney. In particular, the Quantum Optics group
at ETH is gratefully acknowledged for our very pleasant and stimulating collaboration together.
This work has been supported by the Deutsche Forschungsgemeinschaft (DFG)
through the Collaborative Research Center TR185, project B3.

\section*{References}

\bibliography{CRAS_Review_biblio}

\end{document}